# Twofold van Hove singularity and origin of charge order in topological kagome superconductor CsV$_3$Sb$_5$

Mingu Kang[1,2,†], Shiang Fang[3], Jeong-Kyu Kim[2,4], Brenden R. Ortiz[5], Sae Hee Ryu[6], Jimin Kim[4], Jonggyu Yoo[2,4], Giorgio Sangiovanni[7], Domenico Di Sante[8,9], Byeong-Gyu Park[10], Chris Jozwiak[6], Aaron Bostwick[6], Eli Rotenberg[6], Efthimios Kaxiras[11], Stephen D. Wilson[5], Jae-Hoon Park[2,4,†] & Riccardo Comin[1,†]

[1]Department of Physics, Massachusetts Institute of Technology, Cambridge, Massachusetts 02139, USA.

[2]Max Planck POSTECH Korea Research Initiative, Center for Complex Phase of Materials, Pohang 790-784, Republic of Korea.

[3]Department of Physics and Astronomy, Center for Materials Theory, Rutgers University, Piscataway, New Jersey 08854, USA.

[4]Department of Physics, Pohang University of Science and Technology, Pohang 790-784, Republic of Korea.

[5]Materials Department and California Nanosystems Institute, University of California Santa Barbara, Santa Barbara, California 93106, USA.

[6]Advanced Light Source, E. O. Lawrence Berkeley National Laboratory, Berkeley, California 94720, USA.

[7]Institut für Theoretische Physik und Astrophysik and Würzburg-Dresden Cluster of Excellence ct.qmat, Universität Würzburg, 97074 Würzburg, Germany.

[8]Department of Physics and Astronomy, Alma Mater Studiorum, University of Bologna, 40127 Bologna, Italy.

[9]Center for Computational Quantum Physics, Flatiron Institute, 162 5th Avenue, New York, New York 10010, USA.

[10]Pohang Accelerator Laboratory, Pohang University of Science and Technology, Pohang 790-784, Republic of Korea.

[11]Department of Physics, Harvard University, Cambridge, Massachusetts 02139, USA.

Correspondence should be addressed to [†]iordia@mit.edu, [†]jhp@postech.ac.kr, [†]rcomin@mit.edu

**The layered vanadium antimonides $A$V$_3$Sb$_5$ ($A$ = K, Rb, Cs) are a recently discovered family of topological kagome metals with a rich phenomenology of strongly correlated electronic phases including charge order and superconductivity. Understanding how the singularities inherent to the kagome electronic structure are linked to the observed many-body phases is a topic of great interest and relevance. Here, we combine angle-resolved photoemission spectroscopy and density functional theory to reveal multiple kagome-derived van Hove singularities (vHs) coexisting near the Fermi level of CsV$_3$Sb$_5$ and analyze their contribution to electronic symmetry breaking. Intriguingly, the vHs in CsV$_3$Sb$_5$ have two distinct flavors – *p*-type and *m*-type – which originate from their *pure* and *mixed* sublattice characters, respectively. This twofold vHs is unique property of the kagome lattice, and its flavor critically determines the pairing symmetry and ground states emerging in $A$V$_3$Sb$_5$ series. We establish that, among the multiple vHs in CsV$_3$Sb$_5$, the *m*-type vHs of the $d_{xz}$/$d_{yz}$ kagome band and the *p*-type vHs of the $d_{xy}$/$d_{x2-y2}$ kagome band cross the Fermi level to set the stage for electronic symmetry breaking. The former band exhibits pronounced Fermi surface nesting, while the latter contributes via higher-order vHs. Our work reveals the essential role of kagome-derived vHs for the collective phenomena realized in the $A$V$_3$Sb$_5$ family, paving the way to a deeper understanding of strongly correlated topological kagome systems.**

The combination of nontrivial topology and strong electronic correlations engenders new types of topological phenomena and electronic excitations, such as the Majorana fermions in topological superconductors.[1] The kagome lattice, a two-dimensional network of corner sharing triangles (Fig. 1a), has recently emerged as a new frontier in this field owing to its unique electronic band structure comprising Dirac fermions at the Brillouin zone (BZ) corner ($K$), vHs at the BZ edge ($M$), and flat bands across the whole BZ (Fig. 1d). On the one hand, both the linear band crossing at K and the quadratic band touching point at $\Gamma$ become gapped once spin-orbit coupling is included (dotted-lines in Fig. 1d), realizing a singular source of Berry curvature and *nontrivial topology*.[2–4] On the other hand, the diverging density of states at the vHs and flat band can facilitate the emergence of *correlated many-body ground states*.[5–10]

During the past few years, the topological physics of the kagome lattice has been intensively explored and successfully realized in a series of 3$d$ transition metal-based kagome compounds. Notable examples include massive Dirac fermions and intrinsic anomalous Hall conductivity in $Fe_3Sn_2$ and $TbMn_6Sn_6$ (Ref.[11,12]); topological flat bands in (Fe,Co)Sn (Ref.[13–15]); and magnetic Weyl fermions and chiral anomaly in $Mn_3(Sn,Ge)$ and $Co_3Sn_2S_2$ (Ref.[16–19]). In contrast, kagome compounds hosting many-body ground states have been elusive, partly because all previously studied Mn-, Fe- and Co-based kagome metals have their vHs and flat bands located away from the Fermi energy.

In this context, special attention should be given to the recent discovery of ternary metals $A$V$_3$Sb$_5$ ($A$ = K,Rb,Cs) with a vanadium kagome net (Fig. 1a).[20–23] In addition to having a quasi-two dimensional crystal structure and nonzero Z$_2$ topological invariants, these compounds display a collection of correlated ground states including 2×2 inverse star-of-David charge order (Fig. 1b) below $T_{CO}$ = 78 ~ 102 K, and superconductivity below $T_c$ = 0.92 ~ 2.5 K.[20–23] Subsequent investigations uncovered the unconventional character of both phases: a scanning tunneling microscopy (STM) study has revealed that the charge order in $A$V$_3$Sb$_5$ is chiral and induces anomalous Hall conductivity by breaking time-reversal symmetry without magnetism;[24–26] while Josephson STM and ultra-low temperature thermal conductivity experiments suggested that the superconductivity is in a strong-coupling regime with $2\Delta_{sc}/k_BT_c \approx 5.2$ and nodal gap structure.[27,28] High-pressure measurements further elucidated the competition between the charge order and superconductivity by reporting an intriguing double superconducting dome in the *T-P* phase diagram.[28–31] Rotational symmetry breaking (nematicity)[32] and a four-unit-cell periodic stripe phase were also detected below 60 K in CsV$_3$Sb$_5$ and found to coexist with superconductivity.[27,33] All these phenomena draw surprising and powerful parallels with other strongly correlated materials including cuprates and Fe-based high-$T_c$ superconductors. Thus, the $A$V$_3$Sb$_5$ series represents an exciting new platform to study the unconventional correlated physics emerging from the itinerant kagome lattice electrons.

A central, unresolved question in this field is the extent to which the singular features of the electronic structure of kagome lattice are responsible for the rich phenomenology observed in $A$V$_3$Sb$_5$. Intriguingly, in an ideal kagome lattice, earlier theoretical studies have predicted that both chiral density wave and unconventional superconductivity may naturally arise from the kagome-derived vHs at filling fractions $n = 5/12$ and $n = 3/12$ (Fig. 1d).[6,7] At this filling fraction, the Fermi surface and vHs at the M points are perfectly nested (Fig. 1e,f) by a reciprocal vector $\boldsymbol{Q} = (\pi, 0)$ consistent with the 2×2 charge order (see also the schematics in Supplementary Fig. S1). Notably, due to the intrinsic asymmetry of the kagome band structure across the Dirac point (Fig. 1d), the two vHs of the kagome lattice at $n = 5/12$ and $n = 3/12$ are fundamentally distinct.[8] In particular, the Fermi surfaces at the two vHs filling fractions are characterized by two different flavors, *p*-type and *m*-type, depending on the *pure* and *mixed* sublattice characters at the vHs (see Fig. 1e,f for illustration). Such sublattice decoration at the vHs filling is critical to understand the unconventional many-body phases emerging from the kagome lattice as it determines the momentum dependence of the electronic susceptibility and the electron-electron (electron-hole) pairing symmetries of the superconducting (charge ordered) state.[34] The physics of twofold vHs is unique to the kagome lattice, while it is absent in the hexagonal lattice where the two vHs at $n = 3/8$ and $5/8$ filling fractions are related by particle-hole symmetry and therefore identical.

Here we present a high-resolution angle-resolved photoemission spectroscopy (ARPES) study of CsV$_3$Sb$_5$ to address the following key questions: (1) whether the singularity of the kagome band structure is truly responsible for the observed many-body phases in $A$V$_3$Sb$_5$; and (2) whether we can understand the unconventional character of charge order and superconductivity in the $A$V$_3$Sb$_5$ series by identifying the vHs flavor (*p*-type or *m*-type) that contributes to electronic symmetry breaking. Using comprehensive momentum- [($k_x$, $k_y$, $k_z$)] and temperature-dependent datasets, we disentangle the intricate electronic structure of CsV$_3$Sb$_5$ to an unprecedented level of detail and identify multiple *p*- and *m*-type vHs arising from the V *3d* orbital and parity degrees of freedom. Among the multiple vHs identified, we find that the *m*-type vHs from the odd parity

$d_{xz}/d_{yz}$ kagome band contributes to the formation of charge order by promoting Fermi surface nesting, while the *p*-type vHs from the $d_{xy}/d_{x2-y2}$ kagome band contributes via higher-order vHs and enhanced density of states. Our work unambiguously reveals the band-, energy-, and momentum-resolved seeds of charge order in $A$V$_3$Sb$_5$ and lays the groundwork to understand the unconventional many-body phases in topological kagome metals based on the unique vHs physics of the kagome lattice.

We elected to investigate CsV$_3$Sb$_5$ as it is expected to exhibit the most prominently two-dimensional electronic structure [according to resistivity anisotropy data[20] and density functional theory (DFT) calculations] and to be least affected by disorder in the alkali sublattice among all $A$V$_3$Sb$_5$ compounds. In Fig. 2, we present the detailed experimental electronic structure of CsV$_3$Sb$_5$, while the band structure calculated from DFT and its orbital projections are shown in Fig. 1g-j for comparison. As shown in Fig. 2a-d, the Fermi surface of CsV$_3$Sb$_5$ as well as the constant energy contours at E = –0.15, –0.30, and –0.45 eV display the characteristic hexagonal symmetry and BZ expected from the underlying kagome lattice. In CsV$_3$Sb$_5$, four dispersive bands cross the Fermi level. First, we identify a circular electron pocket at the BZ center $\bar{\Gamma}$, which we label as the *G*-band (Fig. 2a,e). The orbital projection from DFT reveals that the *G*-band has dominant Sb characters (see Supplementary Fig. S2). The vanadium kagome net mainly contributes to the Fermi contours near the BZ boundary $\bar{M}$ and $\bar{K}$. Focusing on the $\bar{K}$ point, two sets of kagome-derived Dirac bands can be recognized by their linear band crossings at Dirac points located at $E_{DP1}$ = –0.27 ± 0.01 eV and $E_{DP2}$ = –1.08 ± 0.05 eV (Fig. 2f,g). The Dirac velocities are estimated to be $v_{DP1}$ = (3.54 ± 0.3) ×10$^5$ m/s and $v_{DP2}$ = (5.13 ± 0.3) ×10$^5$ m/s. These Dirac bands arise from the $d_{xy}/d_{x2-y2}$ and $d_{xz}/d_{yz}$ vanadium orbitals, respectively (Fig. 1h,i), and we label these bands as *K1* and *K2*. The fourth band (labeled *K2'*) disperses in the opposite direction to the *K1* and *K2* bands (Fig. 2f-h), away from the $\bar{K}$ point as the binding energy increases. Such behavior of the *K2'*-band can also be traced from the constant energy contours in Fig. 2a-d. The *K1*- and *K2*-bands shrink toward the $\bar{K}$ point while the *K2'*-band contour expands away from $\bar{K}$ with increasing binding energy. The *K2'*-band has the same vanadium $d_{xz}/d_{yz}$ orbital character as *K2* (Fig. 1i) but different parity (see below). We also refer to Supplementary Fig. S3 for ARPES spectra along other high-symmetry directions and their polarization dependence.

It is worthwhile to note that the multiple sets of V-*3d* kagome bands play a crucial role for the topological character of $A$V$_3$Sb$_5$. Due to the opposite dispersion direction, the *K2'*-band intersects the *K1*- and *K2*-bands at various parts of the BZ to form multiple Dirac nodes, as shown in Fig. 1g.[35] Such crossing nodes are clearly observable in our ARPES spectra as marked in Fig. 2f,h. Upon the inclusion of spin-orbit coupling or a time-reversal symmetry breaking field from the chiral charge order, the Dirac nodes acquire a finite mass gap and contribute to the nontrivial $Z_2$ topology and the anomalous Hall conductivity in $A$V$_3$Sb$_5$.[21,24] We also note that the nontrivial π Berry phase associated with the Dirac nodal crossing between *K2* and *K2'*-bands (Fig. 2f) was recently demonstrated in a quantum transport experiment.[36]

With the electronic structure of CsV$_3$Sb$_5$ resolved in full detail, we now focus on the saddle point behavior of the vHs arising from the *K1*, *K2*, and *K2'*-bands near the $\overline{M}$-point. Previous theoretical investigations have proposed that the diverging density of states of the vHs near the Fermi level might be the driving force of charge order and superconductivity in $A$V$_3$Sb$_5$.[35,37] From the high-resolution ARPES spectra of CsV$_3$Sb$_5$ (Figure 3), multiple vHs can be directly identified from their saddle points dispersion with opposite concavities along two orthogonal directions (Fig. 3h,i). For example, in the case of the *K2*-band marked in Fig. 3a, the band bottom of the upper Dirac cone at $\overline{K}$ (Fig. 3a1) moves up in energy upon approaching the $\overline{M}$ point (Figs. 3a2-a5) forming a hole-like band top at $\overline{M}$ along the $\overline{K}$-$\overline{M}$-$\overline{K}$ directions (blue-dotted line). Along the $\overline{\Gamma}$-$\overline{M}$-$\overline{\Gamma}$ direction, however, the *K2*-band retains the electron-like dispersion (panel a5) forming a saddle point or vHs at the binding energy $E_{vHs2}$ = –0.15 ± 0.03 eV. The *K1*-band marked in Fig. 3b also displays a similar saddle point behavior albeit with a less pronounced dispersion (see also Fig. 1g, Fig. 2i and Fig. 5). Surprisingly, we observe a different type of saddle point behavior for the *K2'*-band as shown in Fig. 3c. In this case, the band top of the hole-like dispersion at $\overline{K}$ (Fig. 3c1) moves down in energy as it approaches the $\overline{M}$ point (Figs. 3c2-c5). Thus, opposite to the vHs of *K1*- and *K2*-bands, the vHs of *K2'*-band presents an electron-like (hole-like) dispersion along $\overline{K}$-$\overline{M}$-$\overline{K}$ ($\overline{\Gamma}$-$\overline{M}$-$\overline{\Gamma}$). The two types of vHs with opposite concavities are schematized in Fig. 3h,i.

The coexistence of two distinct vHs flavors near the Fermi level can be captured already at the simplest level of a minimal tight-binding model comprising two orbitals only.[34] According to the $D_{6h}$ point group symmetry, the $d_{xz}/d_{yz}$ orbital tight-binding model on the kagome lattice produces two sets of kagome bands with opposite odd and even mirror eigenvalues $M_{\Gamma M}$ as shown in Fig. 3d.[34] Within this simplified description, the *K2'*-band can be understood as the lower Dirac band of the $M_{\Gamma M}$ = odd kagome set with vHs at $n$ = 3/12 fillings, while the *K2*-band would correspond to the upper Dirac band of the $M_{\Gamma M}$ = even kagome set with vHs at $n$ = 5/12 fillings.

Importantly, as emphasized in earlier theoretical works,[8,34] the two vHs of the kagome lattice at $n = 3/12$ and $n = 5/12$ fillings are fundamentally distinct due to the intrinsic electron-hole asymmetry of the kagome bands (Fig. 1d). To identify the vHs flavors, we extracted the sublattice projections of the *ab initio* Bloch wave functions for the *K1*- *K2*- and *K2'*-vHs. As shown in the charge density profiles in Fig. 3e-g, the *K1*- and *K2*-vHs have dominant *A* sublattice character, while both the *B* and *C* sublattices contribute to the *K2'*-vHs. This provides a direct visualization of the *p*-type (*m*-type) nature of the *K1*- and *K2*-vHs (*K2'*-vHs). With neighboring vHs connected by the nesting vector $Q = (\pi, 0)$ having different sublattice characters (Fig. 1e,f), nonlocal Coulomb interactions become important in describing the nesting-driven many-body electronic orders.[5–8,34] Naturally, the leading instabilities and symmetries of the ground state are markedly different for the *p*-type and *m*-type vHs. For example, the former promotes *p*- and $f_{x3-3xy2}$-wave triplet pairing while the latter favors $d+id$-wave singlet and $f_{y3-3yx2}$-wave triplet paring channels.[5–8,34] These results underscore the importance of identifying the dominant vHs flavors in the $A$V$_3$Sb$_5$ family to understand the properties and symmetries of the emergent many-body phases, including the superconducting gap symmetry, which is still under debate.[28,38]

To this end, we analyze the contribution of each band and vHs to the formation of charge order. To start, we discuss the alignment of *K1*-, *K2*-, and *K2'*-band vHs with respect to the Fermi level ($E_F$) of CsV$_3$Sb$_5$. As shown in Fig. 1g, the three experimentally identified vHs are closely reproduced in the DFT calculation (see coral, blue, and red curves). While the *K1*-vHs and *K2'*-vHs disperse in $k_z$ and cross $E_F$ (see M-L line in Fig. 1g), the *K2*-vHs is always located below $E_F$. Such $k_z$-dependence of the vHs is confirmed by photon energy ($E_{ph}$)-dependent measurements (Supplementary Fig. S4, S5 and Supplementary Discussion 1). Figure 3a-c also demonstrates that at intermediate $k_z$ ($k_z \approx 0.4*\pi/c$, $E_{ph} = 88$ eV) the *K1*-vHs and *K2'*-vHs locate very close to the $E_F$, while the *K2*-vHs is further away from $E_F$. Given their proximity to the Fermi energy, we next focus on the detailed spectral features near the *p*-type *K1*-vHs and *m*-type *K2'*-vHs.

In Figure 4, we examine the tendency of *K2'*-band to promote Fermi surface nesting. We note that photoemission matrix element effects manifest dramatically in the $A$V$_3$Sb$_5$ series, as is common for systems with multiple orbital and sublattice degrees of freedom.[13,39] In particular, the photoelectron intensities from *K2*- and *K2'*-bands with opposite mirror parity are enhanced in very different regions of reciprocal space (see Supplementary Fig. S3 for more details on polarization dependence), which greatly helps disentangling multiple coexisting bands near the Fermi level. In the experimental geometry used for Fig. 4a, for example, the matrix element enhances the *K2*-band in the first BZ, while the *K2'*-band is more prominently visible in the second BZ. It is then apparent that the Fermi surface of *K2* is bent inward, deviating from the perfect nesting condition (Fig. 1e,f) and consistent with the vHs of *K2* being located away from $E_F$ (Fig. 3a). Instead, as highlighted in Fig. 4a, the *K2'*-band forms an almost perfectly nested Fermi surface with nesting vector $Q = (\pi, 0)$. The strong nesting of the *K2'*-band as well as the deviation of the *K2*-band can be fully captured from the calculated Fermi surface displayed in Fig. 4b (see also Supplementary Movie 1 for the $k_z$-dependence of the Fermi surface). The nesting structure and the nesting vector $Q = (\pi, 0)$ of the *K2'*-band is comparable to the ideal vHs case in Fig. 1e and is compatible with the observed 2×2 inverse David star charge order in $AV_3Sb_5$ series (see Supplementary Fig. S1). To directly demonstrate the active role of the *K2'*-band, we extract the band- and momentum-resolved charge order gap in Fig. 4c,d. As can be seen from the symmetrized energy distribution curves and the Dynes fit in Fig. 4d, a 20 meV gap opens along the nested Fermi surface of the *K2'*-band but not on the *K2*- and *G*-bands. In Supplementary Fig. S6 and S7, we further confirmed the gap opening on the *K2'*-band by analyzing the leading-edge shift in the unsymmetrized spectra and by detecting the gap signatures in the ARPES energy-momentum maps. Upon increasing the temperature above $T_{CO}$ to 95 K, both the leading-edge shift and the dip in the symmetrized spectra vanish, confirming that they originate from the onset of long-range charge order. The gap opening demonstrates the connection between the *K2'*-band with *m*-type vHs and the onset of charge order in $CsV_3Sb_5$.

At the same time, the *p*-type vHs from the *K1*-band also contributes to the electronic energy lowering in the charge order state as shown in Fig. 5. The dispersion of *K1*-vHs above $T_{CO}$ exhibits a hole-like dispersion along the $\bar{K}$-$\bar{M}$-$\bar{K}$ direction (Fig. 5a). With lowering the temperature below $T_{CO}$ (Fig. 5b-d), the spectral weight from the *K1*-vHs gradually shifts to the higher binding energy, signaling the opening of charge order gap at $E_F$. At a temperature of 6 K, the *K1*-band develops a characteristic 'M-shaped' dispersion with the largest renormalization observed at the $\bar{M}$-point as depicted in Fig. 5e. However, an important difference compared to the *K2'*-band is that the *K1*-band does not produce a well-nested Fermi surface (see Fig. 4b). This fact is closely related to the higher-order nature of the *K1*-vHs where the energy-momentum dispersion involves contributions higher than quadratic (see Supplementary Fig. S8 for more details). In Fig. 5f, we experimentally confirm the higher-order nature of the *K1*-vHs by showing that its dispersion follows more closely a quartic polynomial instead of a quadratic one. The deviation from the quadratic form warps the

Fermi surface from a perfect hexagon (see schematics in Fig. 5g), explaining the absence of Fermi surface nesting in the *K1*-band. Finally, we note that the observed charge order gap opening on the *K1*- and *K2'*-bands are consistent with the unfolded band structure calculation of $CsV_3Sb_5$ in the 2×2 inverse David star phase, as marked by white and yellows arrows in Fig. 5h.

In conclusion, we resolved the multiple kagome-derived vHs near the Fermi level of $CsV_3Sb_5$ and analyzed their contributions to the electronic symmetry breaking that characterizes $AV_3Sb_5$ compounds. We identified three vHs: the *p*-type vHs from *K1*-band and the *m*-type vHs from *K2'*-band cross the Fermi level, while the *p*-type vHs from *K2*-band is located away from it (Fig. 3). For each band, we consider the diverging density of states at the vHs and the Fermi surface nesting condition (with diverging number of nested *k*-points) as two independent factors setting the stage for electronic symmetry breaking. We find that the *K2'*-band contributes to the electronic symmetry breaking via both *m*-type vHs and Fermi surface nesting (Fig. 4), while the *K1*-band contributes via higher-order *p*-type vHs (Fig. 5).

We note that the charge order in $AV_3Sb_5$ kagome series shares many similarities with that of the high-temperature superconducting cuprates, namely: (i) a weak superlattice peak and small entropy release[21,40]; (ii) the occurrence of symmetry breaking in the absence of a soft mode[41,42]; and (iii) the potential connection to instabilities of the parent electronic band structure.[43] The superconductivity in this system might arise from the same vHs and nesting conditions proposed theoretically.[5–8,34] In this context, the multiple kagome-derived vHs observed here might explain the multiple superconducting domes and Lifshitz transition revealed in high-pressure experiments.[30,31] Connecting the vHs sublattice flavors to the pairing symmetries of many-body phases in $AV_3Sb_5$ is a key step to understand their origin and character. The fact that the same V-*3d* kagome bands also contribute to the topological electronic structure suggests that the many-body ground state and nontrivial topology are closely entwined in the $AV_3Sb_5$ kagome series, thereby establishing an exciting new platform to search for novel composite quasiparticles.

## Methods

**Sample synthesis and angle-resolved photoemission spectroscopy**

High-quality single crystals of CsV$_3$Sb$_5$ were synthesized via flux method as described in Ref.[20,21]. ARPES experiments were performed at Beamline 4A1 of the Pohang Light Source and Beamline 7.0.2 (MAESTRO) of the Advanced Light Source, both equipped with R4000 hemispherical electron analyzer (Scienta Omicron). CsV$_3$Sb$_5$ samples were cleaved inside an ultrahigh vacuum chamber with a base pressure better than $\approx 4 \times 10^{-11}$ torr. For the $k_z$-dependence studies, photon energy was scanned from 70 eV to 200 eV, covering more than three complete three-dimensional Brillouin zone. The main datasets were collected using 95 eV (Fig. 2), 88 eV (Fig. 3), 95 eV (Fig. 4), 113 eV (Fig. 5a-e), and 106 eV (Fig. 5f) photons with linear horizontal polarization. The energy and momentum resolutions were better than 40 meV and 0.01 Å$^{-1}$ for the Pohang Light Source data and better than 20 meV and 0.01 Å$^{-1}$ for the Advanced Light Source data, respectively. We refer to the Supplementary Table 1 for the comprehensive description of the experimental parameters including experimental geometry, analyzer slit direction, sample alignment, and temperature for each data in Fig. 2-5 and Supplementary Fig. S3-8.

**Density functional theory and tight-binding calculations**

DFT calculations were performed using the Vienna Ab initio Simulation Package,[44,45] with GGA-PBE exchange-correlation functional[46] and the pseudo potential formalism based on the Projector Augmented Wave method.[47] We used lattice parameters ($a$, $b$, $c$) = (5.504 Å, 5.504 Å, 9.715 Å), which is determined from a DFT relaxation. The kinetic energy cutoff was 350 eV for the plane-wave basis sets, with a Γ-centered 15×15×11 grid sampling in the Brillouin zone. To simulate the charge ordered state, we consider a 2×2×1 supercell reconstruction with the atomic positions allowed to relax. The crystal structure converged to the stable inverse David star pattern as displayed in Fig. 1b. To visualize the effect of structural deformation to the electronic structure, we unfolded the bands calculated in the ISD state to the original Brillouin zone as shown in Fig. 5h. Spin-orbit coupling terms are not included in the calculation as it only marginally affects the electronic structure near E$_F$ (See Supplementary Fig. S9 for the comparison of DFT band structure with and without spin-orbit coupling).

## Acknowledgements

We acknowledge the fruitful discussions with Prof. Sungwon Jung. This work was supported by the STC Center for Integrated Quantum Materials, NSF Grant No. DMR-1231319. The works at Max Planck POSTECH Korea Research Initiative were supported by the National Research Foundation of Korea, Ministry of Science, Grant No. 2016K1A4A4A01922028. B.R.O. and S.D.W. were supported by the National Science Foundation (NSF) through Enabling Quantum Leap: Convergent Accelerated Discovery Foundries for Quantum Materials Science, Engineering and Information (Q-AMASE-i): Quantum Foundry at UC Santa Barbara (DMR-1906325). This research used resources of the Advanced Light Source, a U.S. DOE Office of Science User Facility under contract no. DE-AC02-05CH11231. M.K. acknowledges a Samsung Scholarship from the Samsung Foundation of Culture. B.R.O. acknowledges support from the California NanoSystems Institute through the Elings Fellowship program. The research leading to these results has received funding from the European Union Horizon 2020 research and innovation program under the Marie Skłodowska-Curie Grant Agreement No. 897276. We are grateful for funding support from the Deutsche Forschungsgemeinschaft (DFG, German Research Foundation) under Germany's Excellence Strategy through the Würzburg-Dresden Cluster of Excellence on Complexity and Topology in Quantum Matter ct.qmat (EXC 2147, Project ID 390858490) as well as through the Collaborative Research Center SFB 1170 ToCoTronics (Project ID 258499086).

## Author contributions

M.K., J.-H.P, and R.C. conceived the project. M.K., J.-K.K, S.H.R., and J.K performed the ARPES experiments and analyzed the resulting data with help from J.Y., C.J., A.B., E.R., and B.-G.P.. S.F., G.S., and D.D.S performed the theoretical calculations with help from E.K. B.R.O. and S.D.W. synthesized and characterized the crystals. M.K., J.-H.P, and R.C. wrote the manuscript with input from all coauthors.

## Competing interests

The authors declare no competing interests.

## Data availability

The data that support the plots within this paper and other findings of this study are available on https://doi.org/10.7910/DVN/KOA1X0.

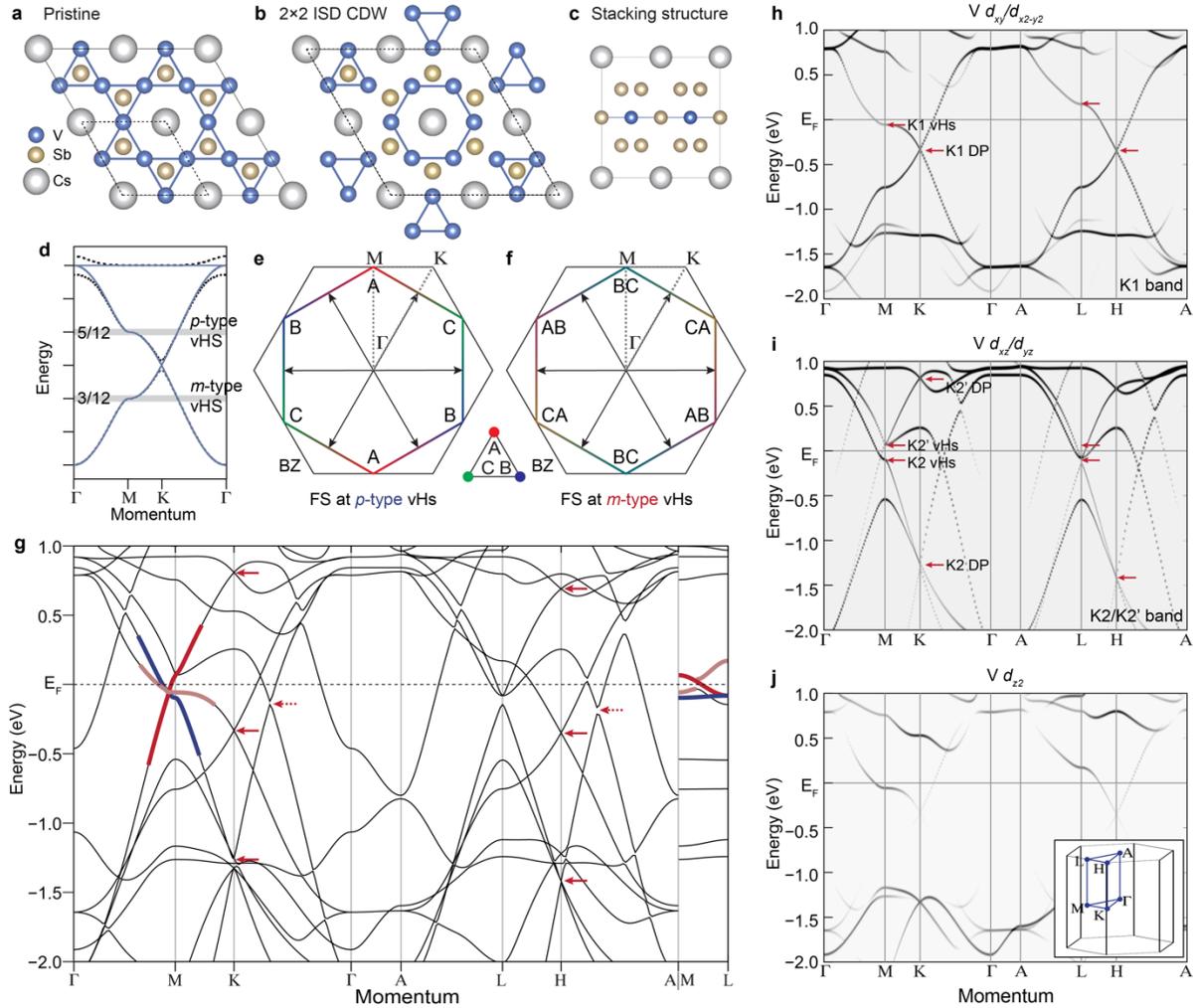

**Figure 1 | Theoretical electronic structure and charge order in kagome metal CsV$_3$Sb$_5$. a,c,** In-plane and out-of-plane lattice structure of CsV$_3$Sb$_5$ with a V-kagome net. **b,** 2×2 inverse David star lattice distortion in the charge order phase of CsV$_3$Sb$_5$. **d,** Prototypical tight-binding electronic structure of the kagome lattice. Dashed- and solid-lines in **c** are the case with and without spin-orbit coupling, respectively. Two van Hove singularities (vHs) at filling fractions $n = 5/12$ and $n = 3/12$ with diverging density of states are marked with grey shades in **d**. **e,f,** Fermi surface of the kagome lattice at the $n = 5/12$ filling with *p*-type vHs and at the $n = 3/12$ filling with *m*-type vHs, respectively. Red, blue, and green colors along the Fermi surface contour represent the distribution of three kagome sublattice weights. The nesting vector $Q = (\pi, 0)$ and its symmetry equivalents are marked with black arrows. **g,** Theoretical electronic structure of CsV$_3$Sb$_5$ from density functional theory. Solid-arrows at K and H mark the multiple Dirac points emerging from the $3d$-orbital degrees of freedom in the V-kagome net, while the dashed-arrows mark the Dirac nodes emerging from the crossing between different kagome sets. Coral, blue, and red solid-lines indicate three saddle-like dispersions or vHs near the Fermi level and their $k_z$-dependences along M-L line. **h-j,** Orbital-projected electronic structure of CsV$_3$Sb$_5$. Th inset in **j** shows the three-dimensional Brillouin zone.

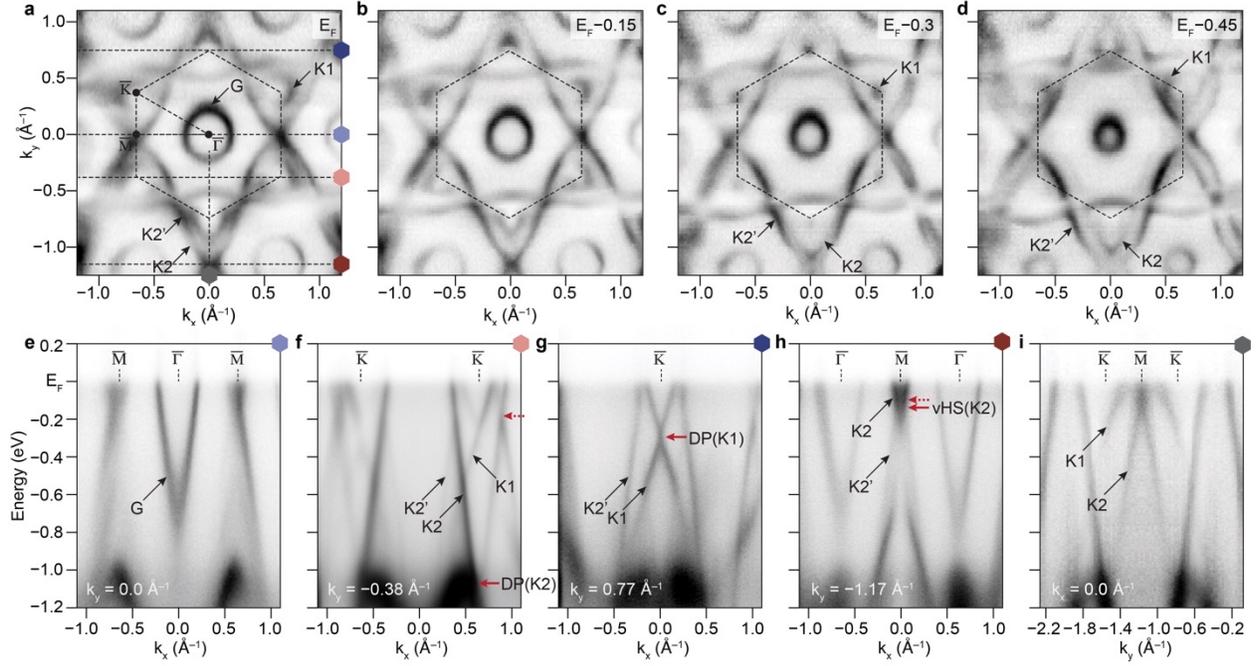

**Figure 2 | Experimental electronic band structure of CsV$_3$Sb$_5$. a-d,** Fermi surface (a) and constant energy contours at E$_B$ = –0.15, –0.30, and –0.45 eV (b-d) as measured with 95 eV photons. The hexagonal surface Brillouin zones are marked with dashed-hexagons in a-d. **e-i,** Energy-momentum dispersion of CsV$_3$Sb$_5$ measured at $k_y$ = 0.0 Å$^{-1}$ (e), –0.38 Å$^{-1}$ (f), 0.77 Å$^{-1}$ (g), –1.17 Å$^{-1}$ (h), and $k_x$ = 0.0 Å$^{-1}$ (i) of a (see also dashed lines in a). The data in i is symmetrized with respect to $\overline{M}$ point at $k_y$ = –1.17 Å$^{-1}$. The four dispersive bands crossing the Fermi level are marked with *G*, *K1*, *K2*, and *K2'*. Solid-red arrows in **e-i** mark the vHs and Dirac points from *K1*- and *K2*-bands while the dashed arrows mark the Dirac nodes emerging from the crossings between *K1*-, *K2*-, and *K2'*-bands. The data in i is symmetrized with respect to $\overline{M}$.

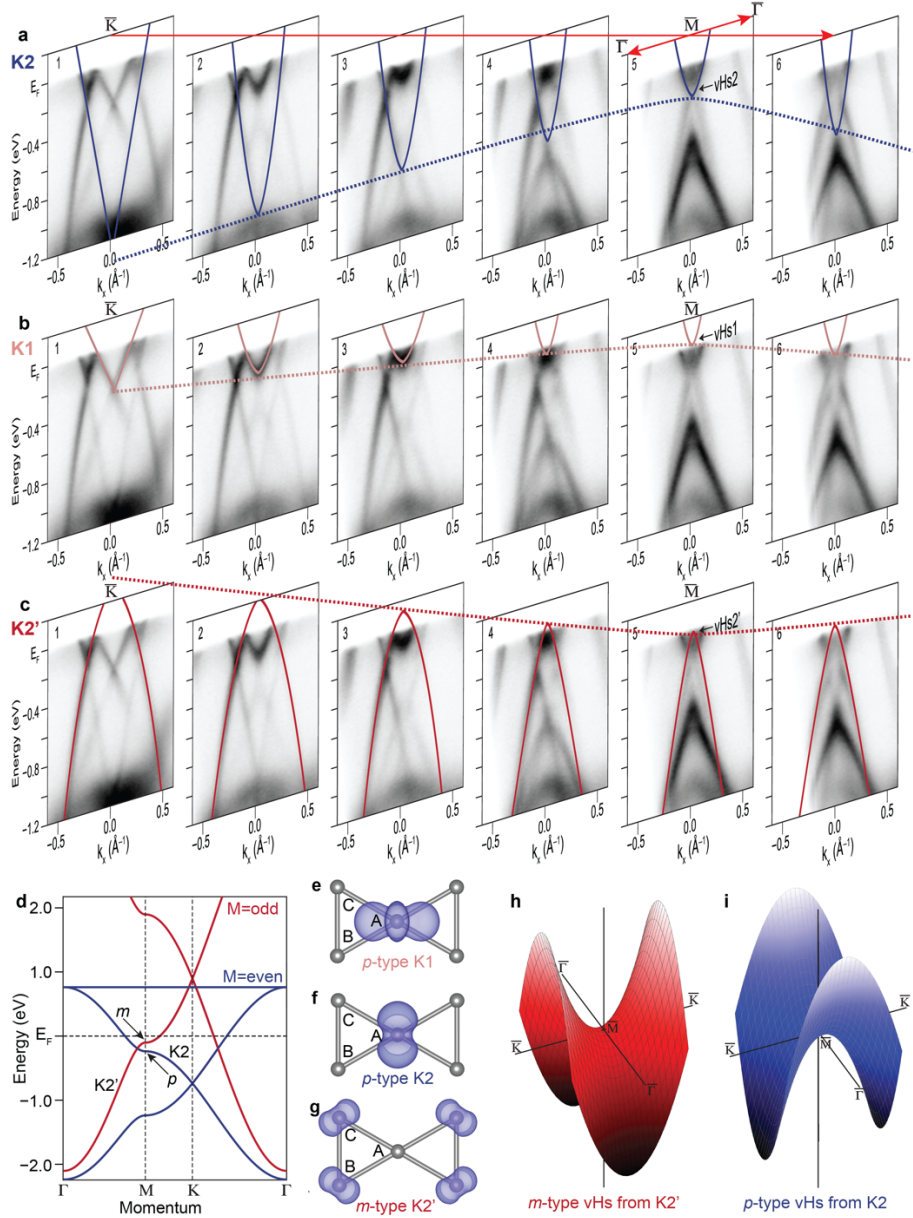

**Figure 3 | Mapping multiple van Hove singularities in CsV$_3$Sb$_5$. a-c,** Tomographic sections of the vHs of *K2*-, *K1*-, and *K2'*-bands. Panels 1-6 correspond to the energy-momentum slices of Fig. 2a at $k_y = -0.76, -0.86, -0.96, -1.06, -1.16,$ and $-1.26$ Å$^{-1}$, respectively. Panels 1 and 5 respectively cross the $\overline{\text{K}}$- and $\overline{\text{M}}$- high-symmetry points. The vHs of *K1*- and *K2*-bands have hole-like dispersion along $\overline{\text{K}}$-$\overline{\text{M}}$-$\overline{\text{K}}$ (dashed-lines in a,b) and electron-like dispersion along $\overline{\Gamma}$-$\overline{\text{M}}$-$\overline{\Gamma}$ (solid-lines in a,b), while the concavity is opposite for the vHs of *K2'*-band (c). The data are obtained with a photon energy of 88 eV, which translates to $k_z \approx 0.4*\pi/c$. **d,** Minimal tight-binding model with $d_{xz}/d_{yz}$ orbitals to describe the *K2*-band with *p*-type vHs and the *K2'*-band with *m*-type vHs. **e-g,** Charge density profiles of the Bloch wave functions at the M-point forming the *K1*, *K2* and *K2'*-vHs, respectively (top view of the kagome plane). **h,i,** Schematics of the *m*-type and *p*-type vHs with opposite concavity.

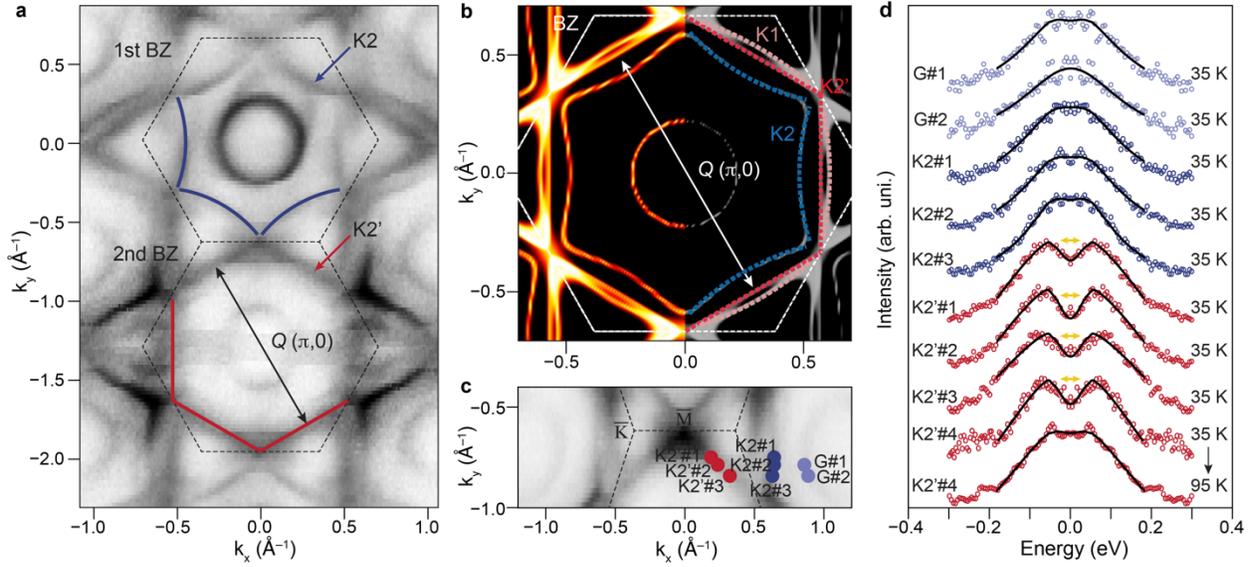

**Figure 4 | Fermi surface nesting and charge order gap in the *K2'*-band. a,** Fermi surface of $CsV_3Sb_5$ across the first and second Brillouin zones. The *K2*- (*K2'*-) band is most intense in the first (second) Brillouin zone as marked by the blue (red) contour. The *K2'*-band displays an almost perfectly nested Fermi surface similar to the simple model of Fig. 1e. **b,** Calculated Fermi energy surface at $k_z = 0.42\ \pi/c$ where the vHs from *K2'*- and *K1*-bands cross $E_F$. Contours of the Fermi surface from *K1*-, *K2*-, and *K2'*-bands are highlighted with coral, blue, and red-dotted lines respectively. Double-headed arrows in a,b indicate the nesting vector compatible with the 2×2 charge order. **c,d,** Band- and momentum-resolved charge order gaps of $CsV_3Sb_5$. **d** shows the symmetrized energy distribution curves as a function of band, momentum, and temperature. Overlaid black-dashed lines represent fits to the Dynes formula. he momentum positions on the Fermi surface from which the energy distribution curves are extracted are marked in b (which are identical data with Fig. 4a).

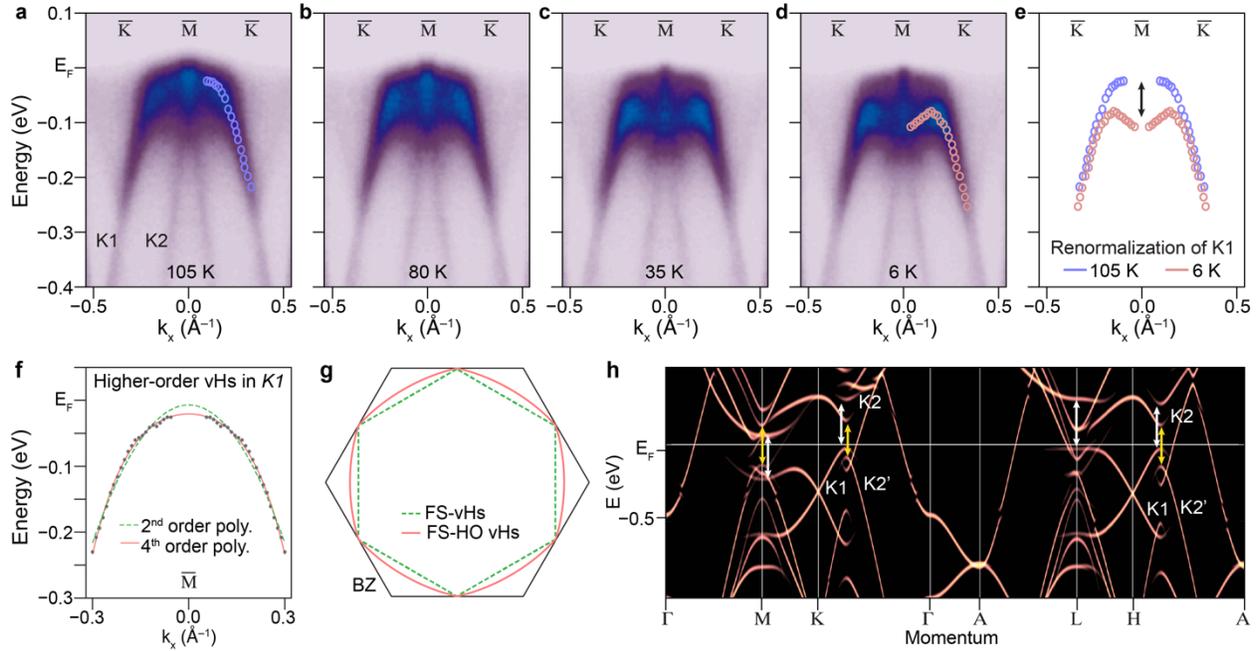

**Figure 5 | Higher-order vHs and charge order gap in the *K1*-band. a-d,** Renormalization of the *K1*-band vHs as a function of temperature measured at 105 K, 80 K, 35 K, and 6 K, respectively. The data is acquired at $k_z \approx$ 0.3 $\pi/c$ (with 113 eV photon energy) where the *K1*-band vHs locates close to the Fermi level. Below $T_{CO}$, the charge order gap opens at the vHs, and the *K1*-band eventually develops the characteristic 'M-shaped' dispersion. **e,** Comparison of the *K1*-band dispersion in the normal (105 K) and charge-ordered state (6 K). **f,** Dispersion of the *K1*-band vHs along $\overline{K}$-$\overline{M}$-$\overline{K}$ direction measured at $k_z \approx 0$ (with 106 eV photons). Green-dotted and coral-solid lines indicate fits to the 2$^{nd}$ order and 4$^{th}$ order polynomials, respectively. The dispersion deviates from quadratic form confirming the higher-order nature of the *K1*-vHs. **g,** Fermi surface of the conventional vHs with perfect nesting (green-dotted line) and of the higher-order vHs that deviates from the nesting condition (coral-solid line). **h,** Unfolded band structure of $CsV_3Sb_5$ in the inverse David star phase. White- and yellow-arrows represent the charge order gap opened at the Fermi level on *K1*- and *K2'*-bands respectively.

# Supplementary Information for
# "Twofold van Hove singularity and origin of charge order in topological kagome superconductor CsV$_3$Sb$_5$"

Mingu Kang[1,2,†], Shiang Fang[3], Jeong-Kyu Kim[2,4], Brenden R. Ortiz[5], Sae Hee Ryu[6], Jimin Kim[4], Jonggyu Yoo[2,4], Giorgio Sangiovanni[7], Domenico Di Sante[8,9], Byeong-Gyu Park[10], Chris Jozwiak[6], Aaron Bostwick[6], Eli Rotenberg[6], Efthimios Kaxiras[11], Stephen D. Wilson[5], Jae-Hoon Park[2,4,†] & Riccardo Comin[1,†]

[1]Department of Physics, Massachusetts Institute of Technology, Cambridge, Massachusetts 02139, USA.

[2]Max Planck POSTECH Korea Research Initiative, Center for Complex Phase of Materials, Pohang 790-784, Republic of Korea.

[3]Department of Physics and Astronomy, Center for Materials Theory, Rutgers University, Piscataway, New Jersey 08854, USA.

[4]Department of Physics, Pohang University of Science and Technology, Pohang 790-784, Republic of Korea.

[5]Materials Department and California Nanosystems Institute, University of California Santa Barbara, Santa Barbara, California 93106, USA.

[6]Advanced Light Source, E. O. Lawrence Berkeley National Laboratory, Berkeley, California 94720, USA.

[7]Institut für Theoretische Physik und Astrophysik and Würzburg-Dresden Cluster of Excellence ct.qmat, Universität Würzburg, 97074 Würzburg, Germany.

[8]Department of Physics and Astronomy, Alma Mater Studiorum, University of Bologna, 40127 Bologna, Italy.

[9]Center for Computational Quantum Physics, Flatiron Institute, 162 5th Avenue, New York, New York 10010, USA.

[10]Pohang Accelerator Laboratory, Pohang University of Science and Technology, Pohang 790-784, Republic of Korea.

[11]Department of Physics, Harvard University, Cambridge, Massachusetts 02139, USA.

Correspondence should be addressed to [†]iordia@mit.edu, [†]jhp@postech.ac.kr, [†]rcomin@mit.edu

## Contents

1. Photon-energy dependent ARPES studies on CsV$_3$Sb$_5$
2. Supplementary Figures S1-S9
3. Supplementary Table 1: Detailed description of the experimental parameters
4. References

1. **Photon-energy dependent ARPES studies on CsV$_3$Sb$_5$**

Despite the overall two-dimensionality of the crystal structure and transport properties of CsV$_3$Sb$_5$, the electronic bands have some dependence on the out-of-plane momentum $k_z$. These include the electron pocket at the zone center $\bar{\Gamma}$ (*G*-band in the main text) and the van Hove singularities (vHs) at the zone edge $\bar{M}$ (see Fig. 1g). In this context, mapping the electronic structure over the complete three-dimensional Brillouin zone (BZ) is essential for a complete understanding of *A*V$_3$Sb$_5$ series.

To this end, we performed extensive photon-energy ($E_{ph}$)-dependent ARPES studies using $E_{ph}$ ranging from 70 eV to 200 eV, covering more than three entire three-dimensional BZs. We first focus on the photon energy dependence of the *G*-band. As shown in Fig. S4a-c, the *G*-band exhibits pronounced photon energy and $k_z$ dependence with the band top along $k_z$ located around –0.5 eV and the band bottom located around –0.85 eV (see Fig. S4c). By comparing this behavior with the Density Functional Theory (DFT) band structure (see G-A line in Fig. 1g), one can identify the band top as Γ ($k_z = 0$) and the band bottom as A ($k_z = \pi/c$). Indeed, the periodicity of the *G*-band tracks the periodicity of the three-dimensional BZ (Fig. S4c), where the corresponding $k_z$ momentum is calculated by assuming nearly free-electron final state with inner potential 2.5 eV.

According to DFT, there always exists a single electron pocket at $\bar{\Gamma}$ independent of $k_z$. Thus, it is intriguing to observe the appearance of two energy-split electron pockets at $\bar{\Gamma}$ in the ARPES data of Fig. S4e,g. The upper band disperses with $k_z$ and represents the bulk band (orange arrows in Fig. S4e,g). In contrast, the lower band is independent on $k_z$ and merges with the bulk band at $k_z = \pi$. We thus assign the lower band to a surface resonance. We note that other ARPES studies reported the variation of *G*-band as a function of time.[1] The surface resonance state observed here can explain this, owing to its sensitivity to the time-dependent oxidization of dangling Cs atoms on the cleaved surface.

2. Supplementary Figures S1-S9

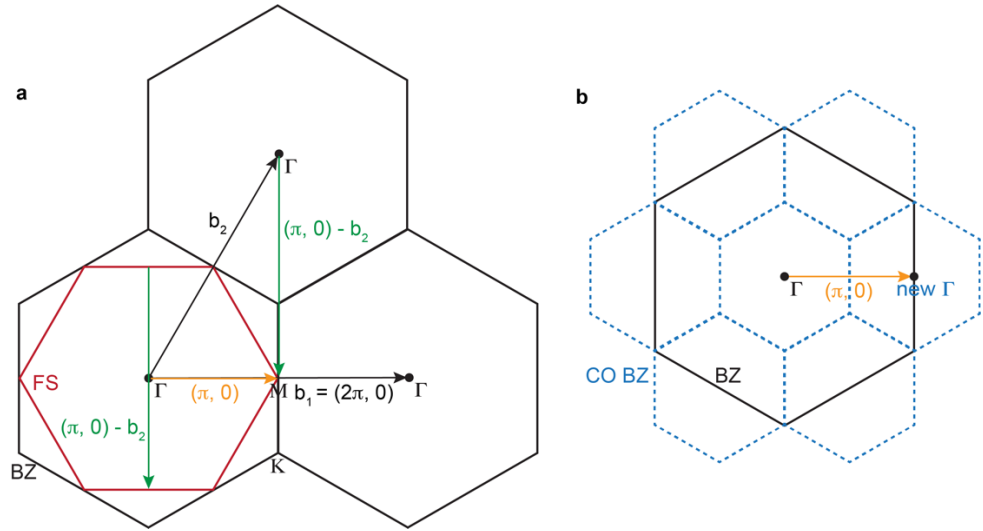

**Figure S1 | Fermi surface nesting and 2 × 2 charge order in the kagome lattice at the vHs filling. a,** Nesting wave vector of the kagome Fermi surface (red hexagon) at the vHs filling. The nesting vector (green arrows) is equivalent to the $(\pi, 0)$ (orange arrow) up to a reciprocal lattice vector (black arrows). **b,** The resulting $(\pi, 0)$ nesting instability folds the original Brillouin zone (black hexagon) to the new Brillouin zone (dotted blue hexagons), in agreement with the expected momentum-space reconstruction induced by 2 × 2 charge order.

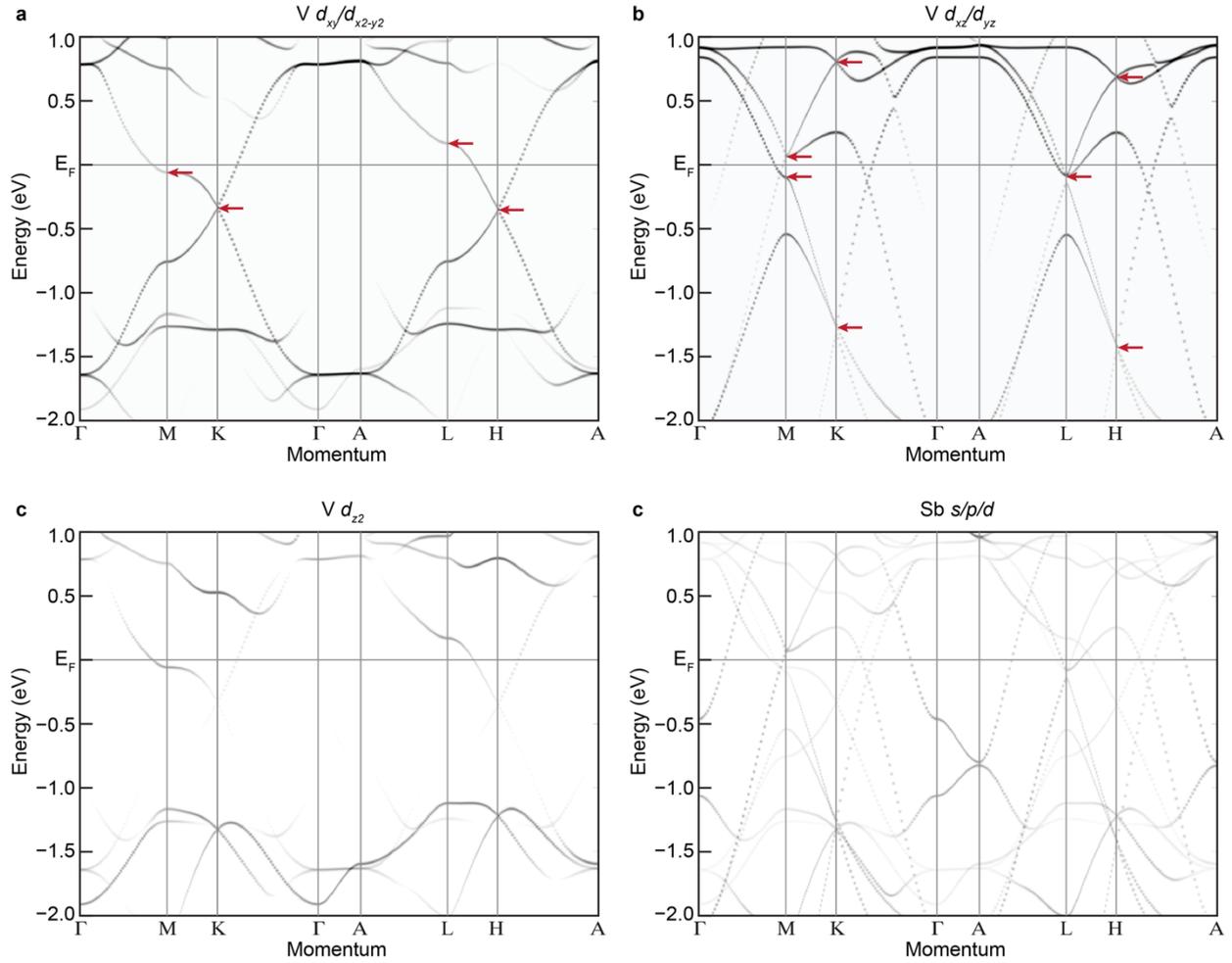

**Figure S2 | Orbital decomposition of the density functional theory (DFT) band structure of CsV$_3$Sb$_5$. a-c,** Orbital weights of vanadium $d_{xy}/d_{x2-y2}$, $d_{xz}/d_{yz}$, and $d_{z2}$ orbitals on the DFT bands of CsV$_3$Sb$_5$, respectively. The kagome-derived van Hove singularities (vHs) and Dirac bands are marked with red arrows in a-c. **d,** Orbital weights of antimony *s*, *p*, and *d* orbitals on the DFT bands of CsV$_3$Sb$_5$.

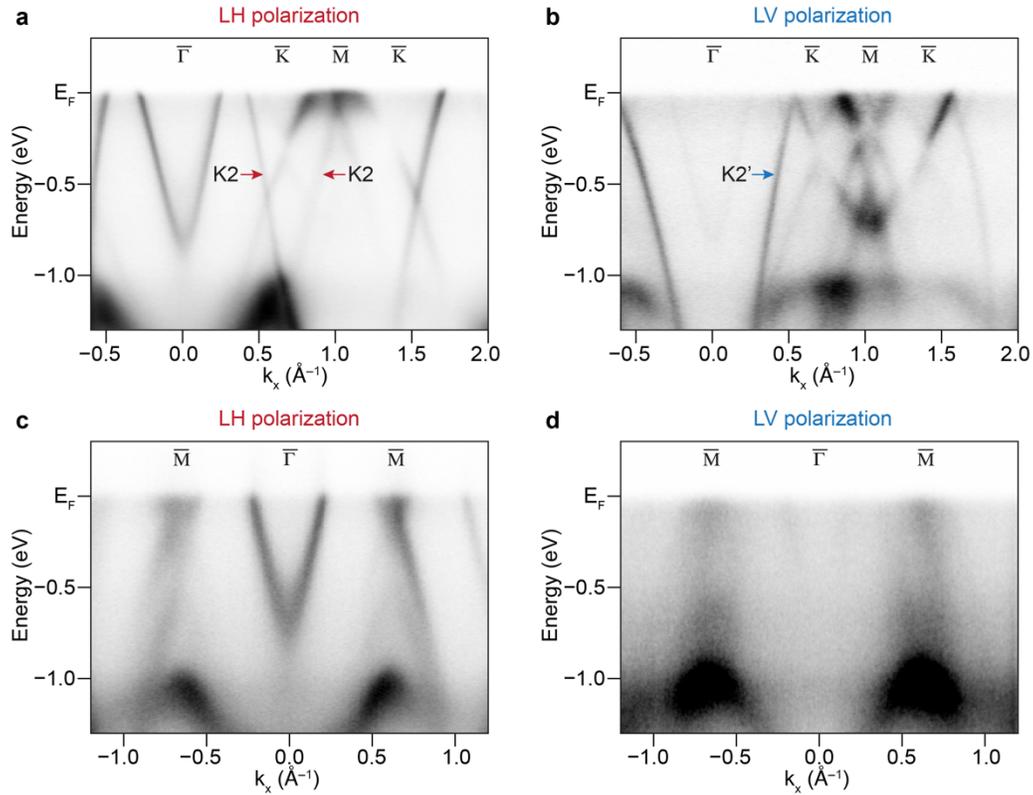

**Figure S3 | Polarization-dependent ARPES spectra in CsV$_3$Sb$_5$ along the $\bar{\Gamma}$-$\bar{K}$ (a,b) and $\bar{\Gamma}$-$\bar{M}$ (c,d) high-symmetry directions.** Data in **a,b** are acquired with 123 eV photons while those in **c,d** are acquired with 92 eV photons. It is noticeable that the *K2*- and *K2'*-bands manifest exclusively in opposite polarization channels: the *K2*- (*K2'*-) band is only visible with linear horizontal (vertical) polarization. The distinct matrix elements for the *K2*- and *K2'*-bands may originate from the opposite mirror parity of the basis orbital (Fig. 3e,f) as discussed in the main text.

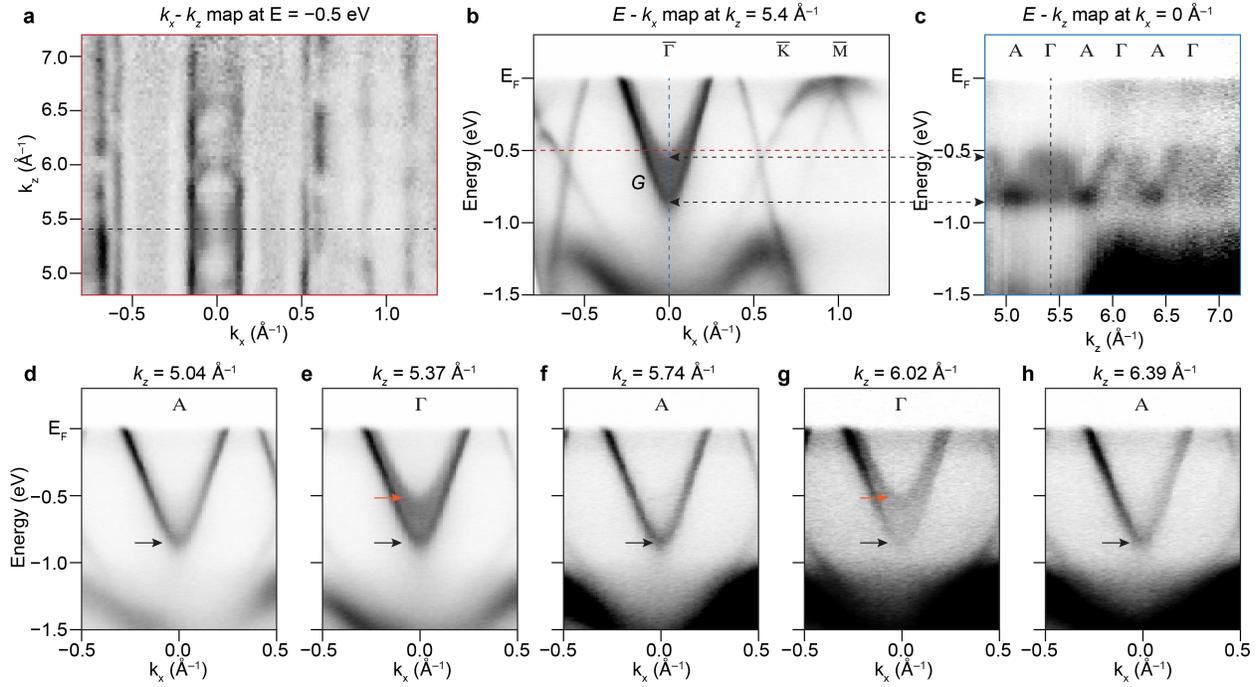

**Figure S4 | Photon-energy dependent ARPES spectra of CsV$_3$Sb$_5$. a,** Representative $k_x$-$k_z$ map of CsV$_3$Sb$_5$ measured at a binding energy –0.5 eV. The data is obtained from wide photon-energy dependent measurements from $E_{ph}$ = 70 eV to 200 eV. **b,** Representative $E$-$k_x$ map of CsV$_3$Sb$_5$ measured at $k_z$ = 5.4 Å$^{-1}$. **c,** Representative $E$-$k_z$ map of CsV$_3$Sb$_5$ measured at $k_x$ = 0 Å$^{-1}$. In c, the energy of the G-band oscillates as a function of $k_z$, with the periodicity following the three-dimensional Brillouin zone of CsV$_3$Sb$_5$. **d-h,** Dispersion of the *G*-band at selected $k_z$ coordinates corresponding to the high-symmetry points Γ (e,g) and A (d,f,h). The orange arrows in e,g indicate the bulk band dispersing in $k_z$, while the black arrows in d-h indicate two-dimensional surface state independent of $k_z$. At $k_z$ = A, the bulk band merges with the surface state. The experimental $k_z$-dependence of *G*-band is highly consistent with the DFT band structure in Fig. 1g.

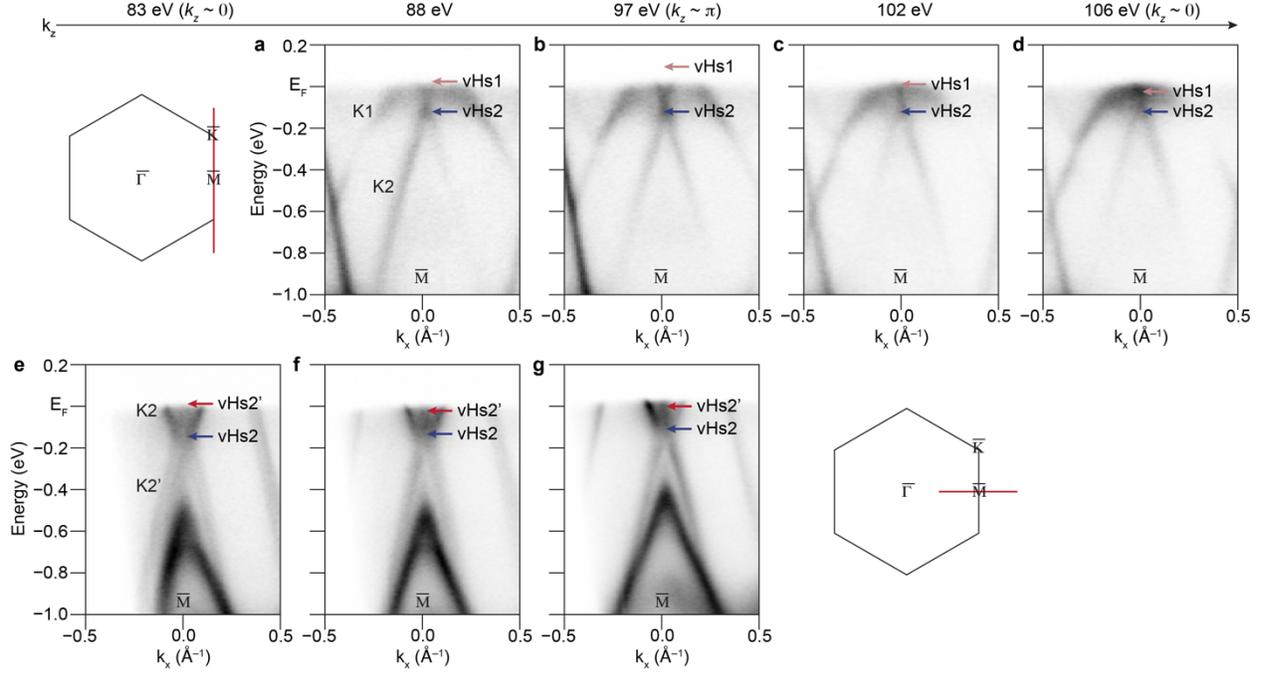

**Figure S5 | $k_z$-dependence of the vHs of *K1*- and *K2*-bands. a-d,** vHs of the *K1*- and *K2*-bands measured along the $\overline{K}$-$\overline{M}$-$\overline{K}$ direction at $E_{ph}$ = 88 eV, 97 eV, 102 eV, and 106 eV respectively. $E_{ph}$ = 97 eV and 106 eV roughly correspond to $k_z \approx \pi/c$ and $k_z \approx 0$, respectively. The vHs from *K1*-band (coral arrow) strongly disperses with $k_z$ and crosses $E_F$ near 102 eV. In contrast, the vHs from *K2*-band (dark blue arrow) is mostly two-dimensional and stays below $E_F$ at all $k_z$. **e-g,** vHs of the *K2'*- and *K2*-bands measured along $\overline{\Gamma}$-$\overline{M}$-$\overline{\Gamma}$ direction at 83 eV, 88 eV, and 97 eV respectively.

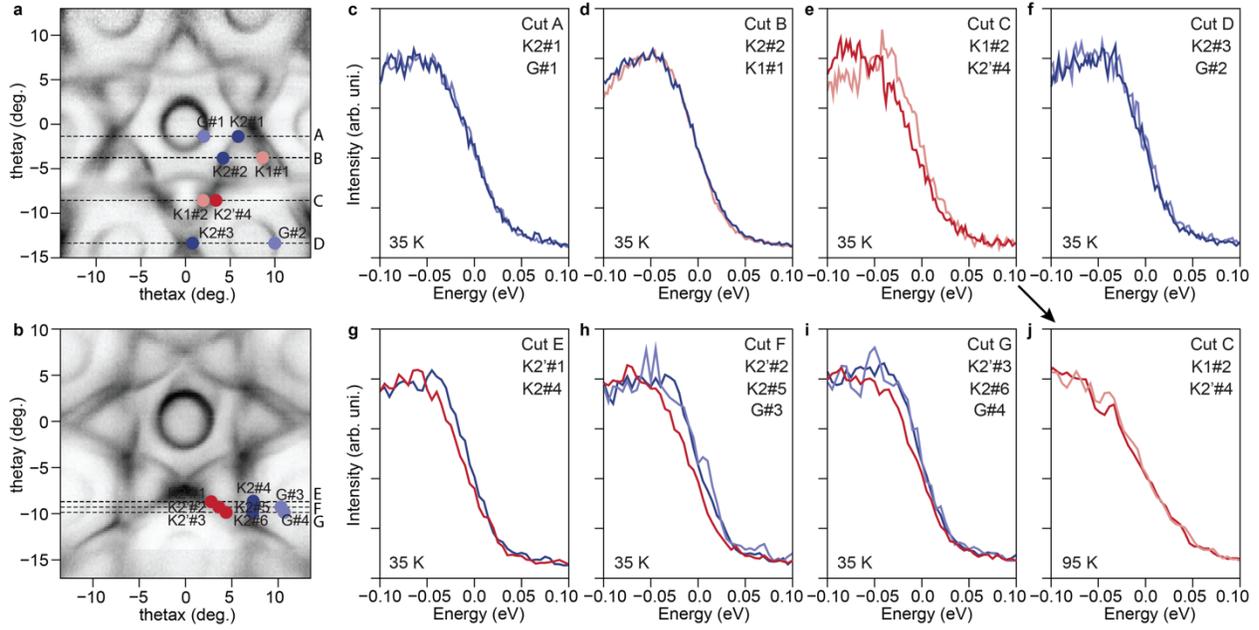

**Figure S6 | Band-resolved charge order gaps and leading-edge shifts in CsV$_3$Sb$_5$. a,b,** Fermi surface of CsV$_3$Sb$_5$ measured in different experimental geometries. The data in a,b are identical to those in Fig. 2a and Fig. 4a respectively. **c-f,g-j,** Energy distribution curves (EDCs) of *G*, *K1*, *K2*, and *K2'*-bands measured at the Fermi momentums marked in a,b. EDCs in c-f, g-i are measured at 35 K (far below $T_{CO}$), while those in j are measured at 95 K (above $T_{CO}$). Appreciable charge order gaps are observed on the *K2'*-band, which exhibits strong Fermi surface nesting and *m*-type vHs close to the Fermi level.

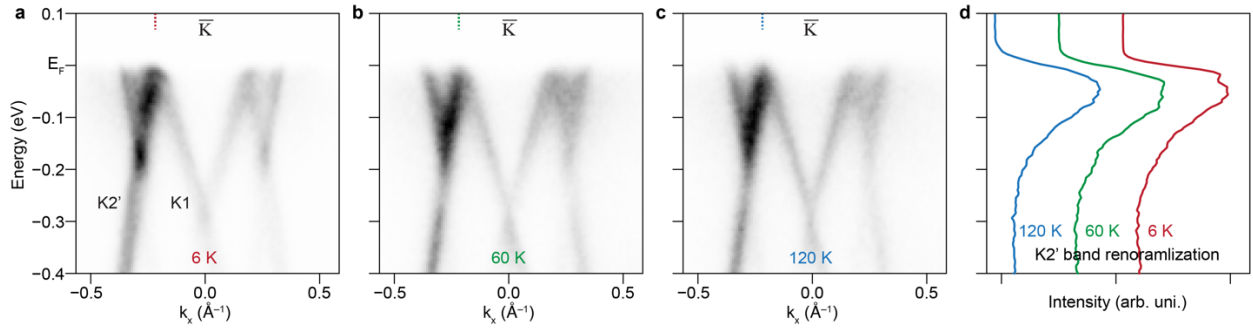

**Figure S7 | Manifestation of charge order gap in the *K2'*-band. a-c,** High-resolution ARPES spectra around the $\bar{K}$ point of $CsV_3Sb_5$ measured at 6 K, 60 K, and 120 K, respectively. Below $T_{CO}$, the spectral weight from the *K2'*-band shifts away from the Fermi level signaling the opening of a charge order gap at the Fermi crossing of the *K2'*-band. **d,** Corresponding energy distribution curves measured at momentum positions marked by dotted lines in a-c.

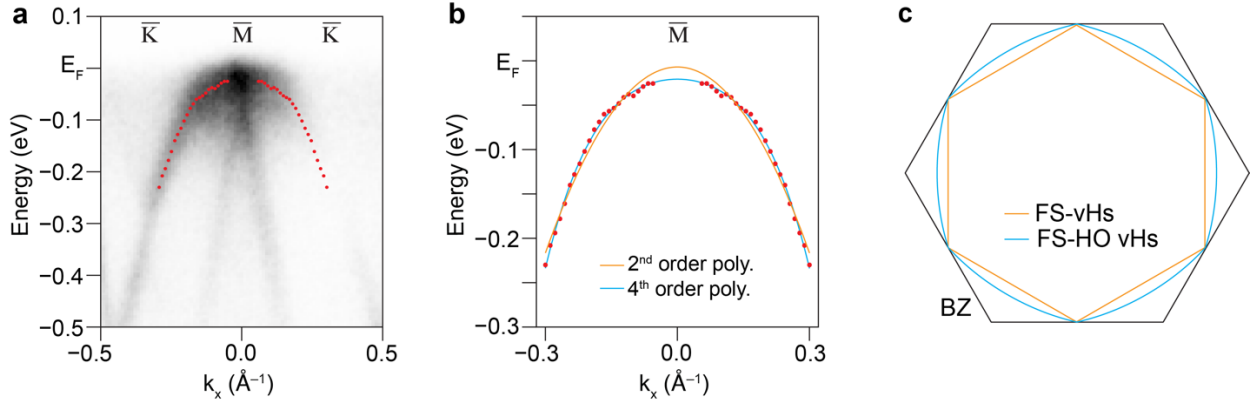

**Figure S8 | Higher-order nature of the *K1*-band vHs. a,** Dispersion of the *K1*-band vHs along the $\overline{\text{K}}$-$\overline{\text{M}}$-$\overline{\text{K}}$ momentum-space direction measured in the normal state (105 K) with 106 eV photons corresponding to $k_z \approx 0$. The dispersion can be precisely tracked from the peak position in the energy distribution curves (overlaid red dots) except very close to $\overline{\text{M}}$ where the intensity from the *K2*-band interferes the analysis. The peak positions are obtained from the negative $k_x$ and symmetrized with respect to $\overline{\text{M}}$. **b,** Polynomial fitting results of the *K1*-band dispersion with quadratic ($E = a_0 + a_2 \Delta k^2$) and quartic ($E = a_0 + a_2 \Delta k^2 + a_4 \Delta k^4$) functions. Fit parameters ($a_0$, $a_2$, $a_4$) = (–0.007 ± 0.002, –2.331 ± 0.046, 0) and (–0.021 ± 0.001, –1.279 ± 0.067, –12.14 ± 0.748) are obtained with energy in eV and momentum in Å$^{-1}$, respectively. A significant contribution from the quartic term is required to fit the dispersion, confirming the higher-order nature of the *K1*-vHs. **c,** Schematics of the Fermi surface with Fermi level at the vHs energy. The orange and blue lines represent the case of conventional and higher-order vHs, respectively.

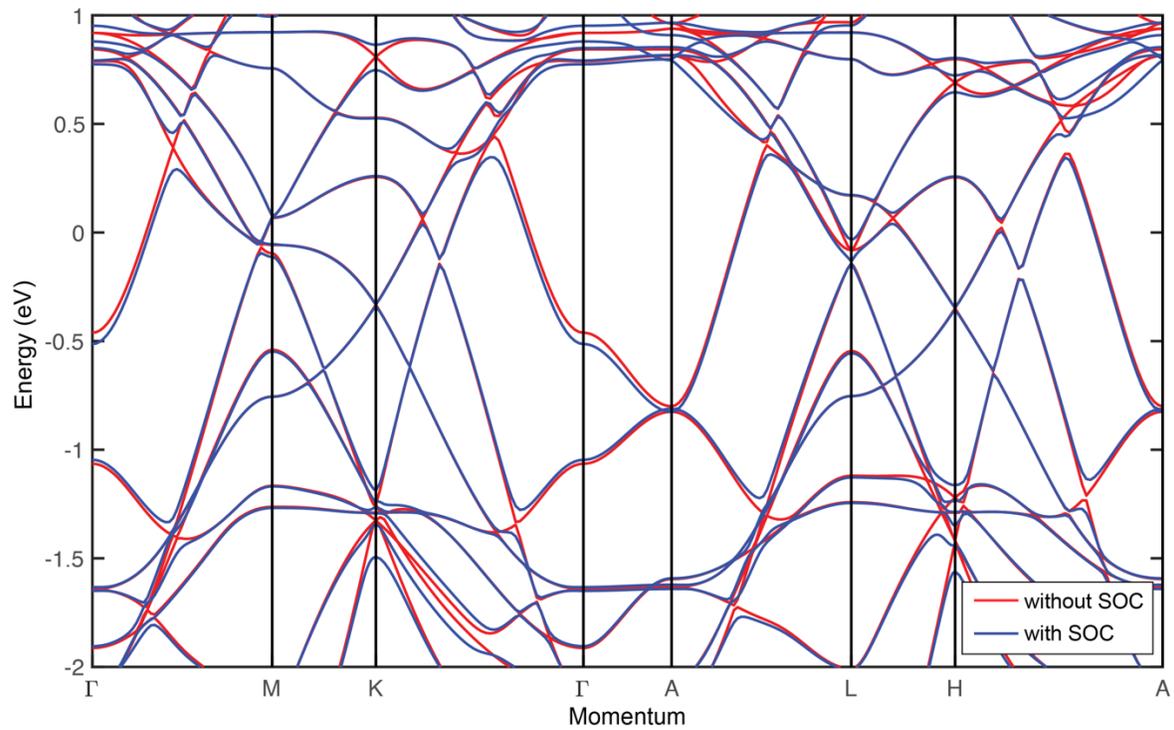

**Figure S9 | DFT band structure of $CsV_3Sb_5$ with and without spin orbit coupling.**

## 3. Supplementary Table 1

|  | Beamline / Synchrotron | Photon energy / Polarization | Sample Temperature | Analyzer slit direction with respect to the scattering plane | Sample orientation with respect to the slit direction |
| --- | --- | --- | --- | --- | --- |
| **Fig. 2** | 4A1 / PLS | 95 eV / LH | 35 K | Vertical | Γ–M |
| **Fig. 3** | 7.0.2 / ALS | 88 eV / LH | 60 K | Horizontal | Γ–M |
| **Fig. 4** | 4A1 / PLS | 95 eV / LH | 35 K, 95 K | Vertical | Γ–K–M |
| **Fig. 5a-e** | 7.0.2 / ALS | 113 eV / LH | 105 - 6 K | Horizontal | Γ–K–M |
| **Fig. 5f** | 7.0.2 / ALS | 106 eV / LH | 105 K | Horizontal | Γ–K–M |
| **Fig. S3a,b** | 7.0.2 / ALS | 123 eV / LH, LV | 105 K | Horizontal | Γ–K–M |
| **Fig. S3c,d** | 7.0.2 / PLS | 92 eV / LH, LV | 85 K | Vertical | Γ–K–M |
| **Fig. S4** | 7.0.2 / ALS | 70 - 120 eV / LH | 105 K | Horizontal | Γ–K–M |
| **Fig. S5** | 7.0.2 / ALS | 83 - 106 eV / LH | 105 K | Horizontal | Γ–M, Γ–K–M |
| **Fig. S6** | 4A1 / PLS | 95 eV / LH | 35 K, 95 K | Vertical | Γ–M, Γ–K–M |
| **Fig. S7** | 7.0.2 / ALS | 88 eV / LH | 6 K - 120 K | Horizontal | Γ–M |
| **Fig. S8** | 7.0.2 / ALS | 106 eV / LH | 105 K | Horizontal | Γ–K–M |

**Table S1 | Detailed description of the experimental parameters.** Acronyms: PLS (Pohang Light Source); ALS (Advanced Light Source); LH (Linear Horizontal Polarization); LV (Linear Vertical Polarization).